\documentclass[conference]{IEEEtran}
\IEEEoverridecommandlockouts
\usepackage{cite}
\usepackage{amsmath,amssymb,amsfonts}
\usepackage{algorithmic}
\usepackage{graphicx}
\usepackage{textcomp}
\usepackage{xcolor}
\usepackage{bbm}
\usepackage{booktabs}
\usepackage{url}
\def\BibTeX{{\rm B\kern-.05em{\sc i\kern-.025em b}\kern-.08em
    T\kern-.1667em\lower.7ex\hbox{E}\kern-.125emX}}
\begin{document}

\title{Efficient Learning of\\Nested Deep Hedging using Multiple Options}

\author{\IEEEauthorblockN{Masanori HIRANO}
\IEEEauthorblockA{
\textit{The University of Tokyo}\\
Tokyo, Japan \\
research@mhirano.jp}
\and
\IEEEauthorblockN{Kentaro IMAJO}
\IEEEauthorblockA{
\textit{Preferred Networks, Inc.}\\
Tokyo, Japan \\
imos@preferred.jp}
\and
\IEEEauthorblockN{Kentaro MINAMI}
\IEEEauthorblockA{
\textit{Preferred Networks, Inc.}\\
Tokyo, Japan \\
minami@preferred.jp}
\and
\IEEEauthorblockN{Takuya SHIMADA}
\IEEEauthorblockA{
\textit{Preferred Networks, Inc.}\\
Tokyo, Japan \\
gct3860+pfn@gmail.com}
}

\maketitle

\begin{abstract}
Deep hedging is a framework for hedging derivatives in the presence of market frictions.
In this study, we focus on the problem of hedging a given target option by using multiple options.
To extend the deep hedging framework to this setting, the options used as hedging instruments also have to be priced during training.
While one might use classical pricing model such as the Black--Scholes formula, ignoring frictions can offer arbitrage opportunities which are undesirable for deep hedging learning.
The goal of this study is to develop a nested deep hedging method.
That is, we develop a fully-deep approach of deep hedging in which the hedging instruments are also priced by deep neural networks that are aware of frictions.
However, since the prices of hedging instruments have to be calculated under many different conditions, the entire learning process can be computationally intractable.
To overcome this problem, we propose an efficient learning method for nested deep hedging. Our method consists of three techniques to circumvent computational intractability, each of which reduces redundant computations during training.
We show through experiments that the Black--Scholes pricing of hedge instruments can admit significant arbitrage opportunities, which are not observed when the pricing is performed by deep hedging.
We also demonstrate that our proposed method successfully reduces the hedging risks compared to a baseline method that does not use options as hedging instruments.
\end{abstract}


\section{Introduction}
Options, a kind of financial derivatives, have become essential tools in modern financial markets.
Options are financial products that ensure rights to buy or sell at predetermined date and price.
Options are traded for many different purposes, such as risk hedging, speculation, and arbitrage \cite{Hull9th}.

The Black--Scholes model \cite{black1973,merton1973} is widely used in practice to calculate options' prices.
However, it has several limitations because of its assumption of the complete market, which is too strong in reality.
In particular, this assumption disregards the existence of \emph{frictions} in the market, such as transaction fees.
Moreover, the normality assumption on the return distribution is also questionable in many real-world market \cite{Cont2001}. 

Deep hedging \cite{deep-hedging} is a recently developed method to overcome the limitations of classical option pricing and hedging methods.
In the deep hedging framework, a hedging strategy is modeled by a neural network, which is trained so that it optimizes the final profit \& loss (PL) under consideration of incomplete markets.
In principle, deep hedging can perform hedging and pricing of arbitrary options, even when their theoretical behaviors are unknown, since deep hedging does not rely on any classical derivative pricing models.

In this study, we develop a deep hedging approach for a problem of hedging a given target option by multiple options.
This setting is of practical importance because of the following reasons.
To reduce the risk of a given option contract, the simplest way is to perform hedging based solely on the underlying stock.
This is similar to a classical hedging method called delta hedging (see e.g., \cite{Hull9th}).
Delta hedging can approximate the price movement of the target option when the price movement of the underlying stock is small.
Delta hedging, however, can fail to reduce the risk when there are significant movements of price or volatility.
Traditionally, the risk of abrupt price changes is controlled by considering other types of higher-order sensitivities, e.g., Gamma and Vega.
Managing these higher-order sensitivities essentially requires trading not only the underlying stock, but also other types of options.
For the case of deep hedging, the situation is quite similar;
If a deep hedging model trades only the underlying stock, it may be still exposed to the risk of abrupt price changes.
Therefore, it is practically important to extend deep hedging to use multiple options.

Ideally, options used as hedging instruments should be priced under the consideration of frictions.
Of course, one can consider an obvious baseline model where the Black--Scholes model calculates prices of hedging instruments without considering frictions.
However, ignoring the effect of friction leads to mis-pricing of hedge instruments, which can offer undesirable arbitrage opportunities to deep hedging.
Therefore, we should handle frictions in a consistent way through the complete learning process.

The above requirement naturally leads to a hierarchical optimization problem described below, which we call \emph{nested deep hedging}.
To calculate the prices of both the hedge instruments and the target options in a consistent way, we consider two types of neural-network-based traders, which we refer to as the primary trader and the secondary traders.
Primary traders optimize the prices of hedge instruments.
The outputs of primary traders, i.e., predicted option prices, are passed into secondary traders aiming to hedge the target option.
Thus, the entire optimization procedure is \textit{nested} in a way that the input variables to the secondary trader have to be optimized in advance by the primary traders.
However, naïve application of deep hedging learning can be computationally intractable.
Since the secondary trader is allowed to trade hedging instruments at every time step until maturity, the prices of the hedging instruments have to be updated many times.
As a result, the primary traders become responsible for calculating the prices of the hedge instruments under many different conditions, thus requiring a large amount of simulations and training iterations.

We propose an efficient learning method for deep hedging to solve the computational problems caused by the nested structure.
Here are our contributions:
\begin{itemize}
  \item We propose an efficient learning method of deep hedging for the situation where an option is hedged by options and stocks.
  \item We show that, by comparing with the deep hedging result, the Black--Scholes model can cause mis-pricing of options under the existence of frictions.
  \item We show that our method performs better in terms of reducing risks, compared to a simpler method that uses solely the underlying stock.
\end{itemize}

\section{Related Work}
\label{sec:related}
The celebrated Black--Scholes model \cite{black1973,merton1973} is certainly the most famous model in the option pricing literature.
To date, the Black--Scholes model has been widely used in practice, whose application include deriving option pricing formulae (e.g., the Black--Scholes formula) and analyzing structures of option prices (e.g., greeks).
Since the Black--Scholes model relies on some restrictive assumptions, its limitation has also been argued.
For example, it assumes that the underlying asset price follows a geometric Brownian motion with a constant volatility, which is pointed out to be unrealistic \cite{Akgiray1989}.
To overcome the limitations, many advances have been made in the literature.
To name but a few, Beckers  {\it et al.} \cite{Beckers1980} introduced the constant elasticity of variance model to relax the normality assumption of returns;
The authors of \cite{Cox1976} and \cite{Merton1976} employed the jump-diffusion models to capture price jumps of underlying assets;
Following this line of work, Madan {\it et al.} \cite{Madan1998} further proposed the Variance-Gamma model.

Advances in neural networks have opened up a new avenue of research on nonparametric approaches to option pricing and hedging.
Hutchinson {\it et al.} \cite{Hutchinson1994} and Garcia {\it et al.} \cite{Garcia2000} showed that multi-layer perceptron (MLP) can replace the Black--Scholes model.
Malliaris {\it et al.} \cite{Malliaris1996} showed that neural networks could predict implied volatility by using the index option of the S\&P100.
For a comprehensive review, see Ruf and Wang \cite{Ruf2020} and references therein.
In their seminal work \cite{Buehler2019,Buehler2019b}, Buehler {\it et al.} proposed a method called \emph{deep hedging}.
Deep hedging constructs option hedging strategies modeled by neural networks, by directly optimizing utility functions defined through the final PL.
Deep hedging has been recognized as a breakthrough in option pricing and hedging because of the two most prominent features:
First, deep hedging can handle various types of frictions by a fully nonparametric approach.
Second, deep hedging does not require any theoretical pricing models, unlike other existing hedging methods such as hedging based on greeks.
Since then, further developments have been made to broaden the applicability of the deep hedging framework.
Horvath {\it et al.} \cite{Horvath2021} proposed a deep hedging model in rough volatility models such as the rBergomi model \cite{Bayer2015}, which includes non-Markov price jumps.
Imaki {\it et al.} \cite{Imaki2021} proposed a deep hedging model considering a no-transaction band \cite{Davis1993} into the neural network architecture and achieved fast convergence of learning.

\section{Task Settings}
As in \cite{Buehler2019}, we will consider a discrete-time optimal hedging problem as follows.
We consider a discrete-time financial market with finite time horizon $T$ and trading dates $0 = t_0 < t_1 < \cdots < t_n = T$.
Suppose there are the following assets:
\begin{itemize}
	\item An underlying stock: $S^{(0)}$
	\item Options used as hedging instruments: $S^{(i)} (i = 1, 2, \cdots, O)$
	\item A target option to be hedged: $S^{(\star)}$
\end{itemize}
We assume that all the hedging instruments $S^{(i)} (i = 0, 1, 2, \cdots, O)$ are \textit{tradable}, while the target option $S^{(\star)}$ is not necessarily tradable.
Here, we say that an asset is tradable if its price is available and can be bought or sold on the market at anytime.
All the options listed above are derived from a single underlying stock, $S^{(0)}$.

\begin{figure}[htbp]
  \centering
  \includegraphics[width=0.7\linewidth]{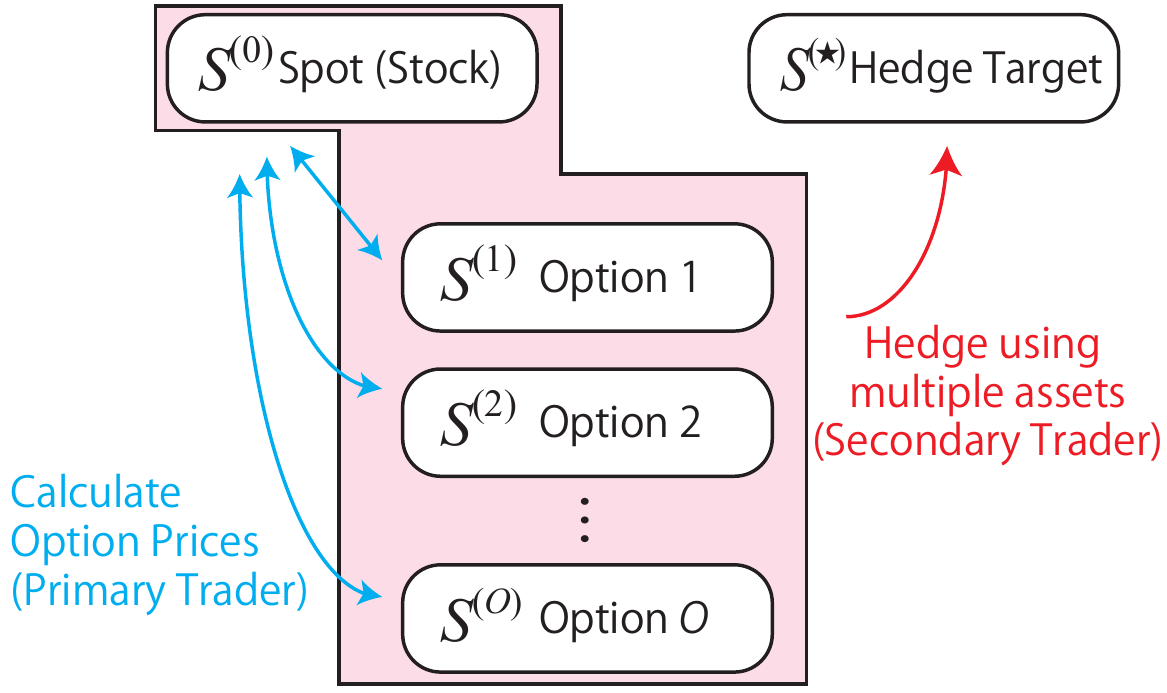}
  \vspace{-3mm}
  \caption{Task setting outline.}
  \label{fig:outline}
\end{figure}

Figure \ref{fig:outline} shows the outline of our task setting.
We consider two types of traders: primary and secondary traders.
Primary traders are responsible for calculating the price of the tradable options  $S^{(i)} (i = 1, 2, \cdots, O)$.
Primary traders try to calculate prices based on the information of the underlying stock prices.
A secondary trader aims to hedge the target option $S^{(\star)}$ using the underlying stock $S^{(0)}$ and all the tradable options.
Here, we assume that the primary traders can always produce appropriate option prices backed by optimal hedging strategies, which eventually ensures that the secondary trader can trade the hedge instruments anytime.

For simplicity, we do not consider risk-free-rate, fix the maturity at $T$, and assume that volatility can be observed directly.

\subsection{Price Process of the Underlying Stock}\label{sec:heston}

As mentioned in Section \ref{sec:related}, an advantage of deep hedging is that it is applicable regardless of the price dynamics of underlying assets.
In the existing literature of deep hedging, widely acknowledged models in finance have been used: for example, the geometric Brownian motion \cite{Buehler2019}, the Heston model \cite{Buehler2019}, and a variant of rough volatility models \cite{Horvath2021}.

In what follows, we focus on the Heston model \cite{Heston1993ACS} defined as
\begin{eqnarray}
	d S^{(0)}_t &=& S^{(0)}_t \sqrt{V_t} dW^{(1)}_t\\
	dV_t &=& \kappa (\theta - V) dt + \sigma \sqrt{V_t} dW^{(2)}_t,
\end{eqnarray}
where $dW^{(1)}_t$ and $dW^{(2)}_t$ are the Wiener process whose correlation coefficient is $\rho$.
In particular, the underlying stock prices $S^{(0)}_{t} ~ (t \in \{ t_0,\cdots, t_n\})$ are generated by some discrete simulation method, such as QE-M Method \cite{andersen2007}.
The Heston model is a sensible choice for our purpose because many asset prices are known to have time-varying volatility, which motivates us to manage the sensitivity to volatility changes (i.e., Vega).

\subsection{Primary Traders: Pricing of the Tradable Options}\label{sec:primary-trader}
Primary traders aim to calculate prices of options used as hedge instruments. Below, we will introduce two types of primary traders: deep hedging and the Black--Scholes formula.

In principle, a price of an option can be determined as the amount of cost required to perform optimal hedging of the option.
On the basis of this idea, we can employ several different pricing methods.
An obvious baseline method is the classical Black--Scholes formula.
The Black--Scholes formula gives a price under the assumption of the complete market, that is, a friction-less market allowing continuous-time trading.
Although a perfect hedge is attainable under the complete market, this is not the case for the discrete-time market with frictions.
In contrast, deep hedging solves optimal hedging problems (and thus calculates the option prices) numerically by neural network learning.
In this sense, deep hedging may perform better in our incomplete market setting.

\subsubsection{\bf Deep Hedging as a Primary Trader}
The prices of tradable options are calculated based on the price of the underlying stock.
More precisely, the price of a tradable option $S^{(i)} (i = 1, \cdots, O)$ at time $t_j$ is calculated by the following procedure:
\begin{enumerate}
  \item[(a)] The primary trader takes a short position of (i.e., sells) a piece of option $S^{(i)}$ at time $t_j$. Note that the price of the option $S_{t_j}^{(i)}$ at this time is still unknown and granted as the placeholder.
  \item[(b)] From the time $t_j$ to the maturity $t_n=T$, the primary trader trades the underlying stock $S^{(0)}$, where proportional transaction costs (with coefficient $c_0$) are imposed on trading actions. 
  \item[(c)] At maturity, the primary trader clears the holding stock and the option based on the strike price.
\end{enumerate}
It is a standard assumption in option pricing theory that the above procedure does not cause any arbitrage opportunities.
In our context, $S_{t_j}^{(i)}$ can be calculated by satisfying the expected total cash flow of the above procedure is $0$.
It can be formulated as:
\begin{align}
    \mathrm{CF} & =
    S^{(i)}_{t_j} + \delta^{(0)}_{t_{n-1}} S^{(0)}_{t_n} - Z_i\left(S^{(i)}\right)\nonumber\\
    &- \sum_{k=j}^{n-1} \left\{
        \left(\delta^{(0)}_{t_k} - \delta^{(0)}_{t_{k-1}}\right)
        S^{(0)}_{t_k}
        + c_0 \left|\delta^{(0)}_{t_k} - \delta^{(0)}_{t_{k-1}}\right| S^{(0)}_{t_k}
    \right\}\label{eq:option_cash_flow} \\
    \mathbb{E}\left[ \mathrm{CF} \right] & = 0 \label{eq:expected_cash_flow}
\end{align}
where $\delta^{(0)}_t$ is the position of the underlying stock $S^{(0)}$ at the time $t$ ($\delta^{(0)}_{t_{j-1}}$ is set as $0$), $c_0$ is the proportional cost for the underlying stock $S^{(0)}$ , and $Z_i\left(S^{(i)}\right)$ is the payoff function of the option $S^{(i)}$.
When the option is European call option and strike is $K^{(i)}$, $Z_i\left(S^{(i)}\right) = \max\left(S^{(i)}_{t_n} - K^{(i)}, 0\right)$.

In deep hedging, the hedging strategy is modeled by a neural network.
In particular, the position $\delta^{(0)}_{t_{k}}$ of the underlying stock is determined as
\begin{eqnarray}
	 \delta^{(0)}_{t_k} = \mathrm{NN}_i\left(I_{t_k}, \delta^{(0)}_{t_{k-1}}\right),
	 \label{eq:NN-underly}
\end{eqnarray}
where $I_t$ is any market information available at time $t$.

To train the neural network model, we minimize the \textit{hedge cost}.
This is equivalent to minimize the negative of the expected cash flow $- \mathbb{E}[\mathrm{CF}]$, where the option price $S_{t_j}^{(i)}$ is ignored as an irrelevant constant during optimization.
After optimization, $S_{t_j}^{(i)}$ is eventually set so as to satisfy \eqref{eq:expected_cash_flow}.
Hence, we can also interpret the entire learning process as finding the minimal option price satisfying the no-arbitrage condition.

Note that the standard formulation of deep hedging \cite{deep-hedging} (and thus that of the secondary trader) involves some notion of utility.
In particular, the secondary trader (introduced in Section \ref{sec:secondary} below) optimizes an objective akin to $\mathbb{E}[u(\mathrm{CF})]$.
However, instead, primary traders optimize the raw total cash flow $\mathbb{E}[\mathrm{CF}]$.
This is because, for our purpose, we do not have to take account of the risk appetite of each individual primary trader, and the roles of primary traders are just to eliminate arbitrage opportunities in the option market.

\subsubsection{\bf Black--Scholes as a Primary Trader}
The Black--Scholes model accepts some market information as inputs and calculates the price of options.
The inputs depend on the option type.
For example, the price calculation of the Black--Scholes model for European options requires the strike, moneyness, time to maturity, and volatility.
Although the implied volatility is used in practice, we used volatility of underlier directly for simplification.

\subsection{Secondary Trader: Hedging Target Option}
\label{sec:secondary}
The secondary trader can trade any tradable assets $S^{(i)} (i = 0, 1, \cdots, O)$, including the underlying stock $S^{(0)}$ and hedge target option $S^{(\star)}$.

First, the secondary trader sells (short) one hedge target option $S^{(\star)}$ at the beginning, and then hedges it to maximize the utility of the hedge.
The total cash flow is:
\begin{align*}
  \mathrm{PL}_T(\delta) &= -Z_{\star}\left(S^{(\star)}\right) + (\delta\cdot S)_T - C_T(S, \delta)\\
  (\delta\cdot S)_T &\equiv \sum_{i=0}^{O} \left\{\delta^{(i)}_{t_{n-1}} S^{(i)}_{t_n} - \sum_{k=0}^{n-1} \left(\delta^{(i)}_{t_k} - \delta^{(i)}_{t_{k-1}}\right) S^{(i)}_{t_k}\right\}\\
  C_T(S, \delta) &\equiv \sum_{i=0}^{O} \sum_{k=0}^{n-1} c_i \left|\delta^{(i)}_{t_k} - \delta^{(i)}_{t_{k-1}}\right| S^{(i)}_{t_k},
\end{align*}
where $Z_{\star}$ is the payoff function of option $S^{(\star)}$, $\delta^{(i)}_{t}$ is the position for the asset $S^{(i)}$ at the end of time-step $t$, $c_i$ is the proportional cost coefficient for the asset $S^{(i)}$, and $\delta$ is the tuple of all asset at the all time, i.e., $\left\{\left.\delta^{(i)}_{t_j} \right| i\in \left\{0, 1, \cdots, O\right\}, j \in \left\{0,\cdots n-1\right\}\right\}$.
Here, $\delta^{(i)}_{t_{-1}}=0$.

The training objective is defined as
\begin{eqnarray}
  J(\delta) = \mathbbm{E}\left[u\left(\mathrm{PL}_T(\delta)\right)\right],\label{eq:second-objective}
\end{eqnarray}
where $u$ is a utility function reflecting the second trader's risk appetite.
For instance, we can set $u$ as (the negative of) some risk measures such as CVaR (Conditional Value at Risk; Expected short-fall) and ERM (entropic risk measure). For precise definitions of these risk measures, see e.g., \cite{deep-hedging}.

As in Section \ref{sec:primary-trader}, the position $\delta$ is modeled by a neural network:
$
  \delta_{t_k} = \mathrm{NN}\left(I_{t_k}, \delta_{t_{k-1}}\right).
$
Here, $\mathrm{NN}$ is a neural network generating $O+1$ outputs of positions, and $I_t$ is any market information available at time $t$.


\section{Proposed Learning Method}
\subsection{Computational Problems in Nested Deep Hedging}
\label{sec:computational_problem}
\begin{figure*}[htbp]
  \centering
  \includegraphics[width=0.8\linewidth]{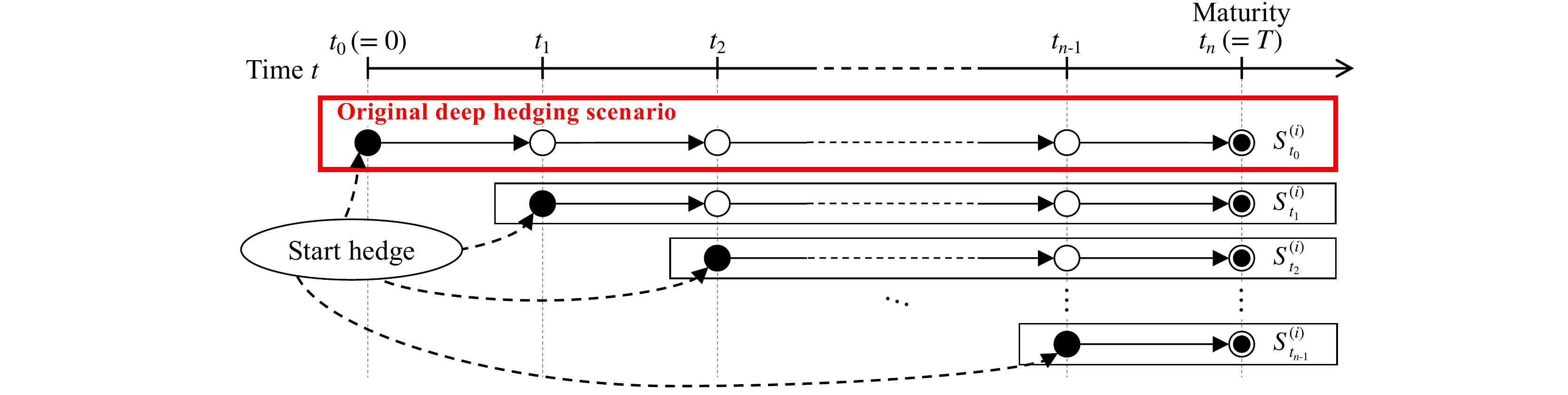}
  \vspace{-3mm}
  \caption{Hedging situations that should be handled by the Deep Hedging model of the primary traders.}
  \label{fig:pricing-ex}
\end{figure*}
As we mentioned in the introduction, the hierarchical optimization problem arising from the nested deep hedging may lead to substantial computational difficulties. Herein, we clarify this point.

The first problem is that the primary traders should handle various hedging situations.
To clarify the issue, we start with recalling the original \textit{non-nested} setting of deep hedging, in which we aim to hedge a single option by trading an underlying stock.
The optimal hedging strategy, and thus the corresponding option price, depends on several parameters such as the time to maturity and the spot price of the underlying stock.
In the non-nested setting, dependence on time to maturity is implicit because they can be considered as fixed during training.
As such, we may train a neural network model specialized to a single time to maturity.

However, in our \textit{nested} setting of deep hedging, the situation is much more complicated because the primary traders have to calculate option prices for many different times to maturity.
Figure \ref{fig:pricing-ex} illustrates the hedging situations that the primary trader should handle.
Since, at each time step $t_i$, the secondary trader requires prices of all the hedge instruments, the primary traders have to update the prices every time with the decrease in the time to maturity.
Moreover, since the training of deep hedging relies on simulations of the underlying stock prices, initial state parameters to be passed to the primary traders vary with time.
For example, if we use the Heston model defined in Section \ref{sec:heston}, the state variables (the spot price and the volatility) are generated randomly and evolve according to the governing equation.
Thus, unlike the non-nested case, the primary traders are responsible for calculating the option prices for a wide range of different combinations of the time to maturity and the initial state parameters.
This causes a difficulty in designing the neural network architecture for the primary traders; If we build a neural network independently for every possible combination of parameters, it will be computationally unrealistic.

The second problem is that the number of simulation paths to ensure the convergence of the training can become very large.
To understand this, we consider a thought experiment where the primary traders are newly trained every time the secondary trader needs to evaluate prices of hedge instruments.
For brevity, we postulate that each primary and secondary trader requires $P$ training examples to ensure convergence.
Then, every time a single simulation path is generated for the training of the secondary traders, we need additional $O(P)$ paths to train the primary traders.
Consequently, the entire training process requires at least $O(P^2)$ calls of the simulator of the underlying stock.
The number of training examples $P$ to train a deep neural network is typically large (say more than 1,000), and a straightforward implementation of the above training procedure could be intractable.

Hence, the overall optimization procedure could be unfeasible due to the aforementioned computational difficulties.
According to our rough estimation, if our experiment were to be implemented naïvely in the sense of the above thought experiment, it would take approximately 1,300 years \footnote{Using $3$ NVIDIA V-100 GPUs in parallel (In the case of $O=2$)} .
In contrast, our proposed method took approximately one day.


\subsection{Proposed Solution for Nested Deep Hedging}
To overcome the aforementioned computational problem, we propose a novel learning scheme for the nested deep hedging.
The following three are the key ingredients of our proposed learning scheme:
(1) Joint training for different times to maturity,
(2) Scheduling of training based on time to maturity,
(3) Optimized simulation for training.

\subsubsection{\bf Joint Training for the Different Times to Maturity}
\label{sec:joint_training}
As we mentioned in the previous subsection, the primary traders should handle various hedging situations that mainly come from different times to maturity.

For this, one might consider reusing a neural network trained over simulation paths with the longest time to maturity (i.e., paths starting from $t = t_0$) in other hedging situations.
However, this approach may not work well in practice, because optimal hedging strategies can be considerably different when times to maturities differ.
For example, it is intuitive that the optimal strategy when the time to maturity is sufficiently long (e.g., three months) can be much different from that for immediate maturity (e.g., tomorrow).

To overcome this issue, we introduce the \textit{joint training} of multiple hedging situations with different times to maturities.
We train $O$ neural networks in total, each of which corresponds to one of the options used as a hedge instrument.
Each network calculates prices of a single option $S^{(i)}$ for all time steps at which we need the price.
When a primary trader aims to calculate the price at time $t_j$, the positions before $t_j$ are forced to be zero, i.e., $\delta_{t_0} = \cdots = \delta_{t_{j-1}} = 0$.
Hence, in our method, we train each network jointly under different constraints on locations of zero positions.



\subsubsection{\bf Scheduling of Training Based on Time to Maturity}

To achieve the joint training described in Section \ref{sec:joint_training}, we here point out that designing scheduling rules of training is important.

First, it is worth noticing that the longer the time to maturity becomes, the more difficult learning optimal hedging strategies becomes.
We can see from \eqref{eq:NN-underly} that deep hedging has the following recurrent structure:
\begin{eqnarray}
  \delta^{(0)}_{t_{n-1}} &=& \mathrm{NN}_i(I_{t_{n-1}}, \delta^{(0)}_{t_{n-2}}) \nonumber\\&=& \mathrm{NN}_i(I_{t_{n-1}},  \mathrm{NN}_i(I_{t_{n-2}}, \delta^{(0)}_{t_{n-3}}))=\cdots. \nonumber
\end{eqnarray}
Here, the recursion will become deeper as the time to maturity is longer.
Learning of deep recurrent networks is generally difficult because of its slow convergence (see e.g., \cite{Imaki2021}).

Second, although an entire trajectory of a hedging strategy consists of multiple actions, optimizing the initial action can be much more important than others.
This comes from the fact that the initial hedging actions (represented as black circles in Figure \ref{fig:pricing-ex}) are qualitatively different from the second or later actions (white circles in Figure \ref{fig:pricing-ex});
At the time of starting hedging (say $t = t_j$), any hedging strategy starts from a zero position (i.e., $\delta_{t_j}^{(0)} = 0$), and thus the primary trader has to trade large amount of stocks in order to neutralize the Delta.
At the subsequent time steps, the primary trader already has a nearly delta-free position, and the updates of the positions are relatively small.

Hence, from the perspective of neural network training, the pricing of $S_t^{(i)}$ is not equally important between time $t = t_0, \ldots, t_{n-1}$.
Therefore, to design an efficient scheduling rule of training, we should prioritize shorter times to maturities.

In our method, we train each primary trader in a \textit{backward} manner with respect to the starting time.
More precisely, the neural network learning (backpropagation + weight update)\footnote{Although we also tried a method in which weight updates are carried out after all backpropagations, it does not work well in convergence and performance.} is sequentially performed from shorter times to maturities.
In other words, in each epoch of learning, we start with training of the hedging strategy for the price at time $t = t_{n-1}$ and continue in descending order $t=t_{n-2}, t_{n-3}, \cdots, 0$.

\subsubsection{\bf Optimized Simulation for Training}
\begin{figure*}[tbp]
  \centering
  \includegraphics[width=0.8\linewidth]{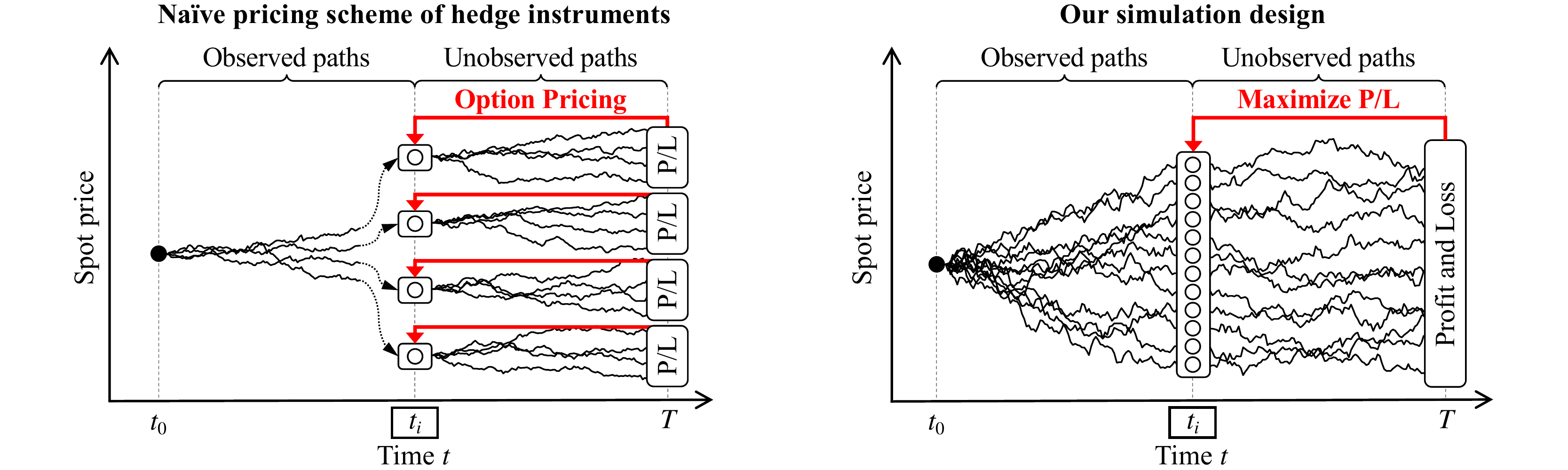}
  \vspace{-3mm}
  \caption{Simulation optimization for training.}
  \label{fig:sim-path}
\end{figure*}
As the third key ingredient of our proposed method, we introduce an optimized simulation design of the underlying stock prices.

As we mentioned in Section \ref{sec:computational_problem}, the number of simulation paths can be quite large.
To illustrate this point, we revisit the na\"{i}ve implementation of the nested deep hedging described in Section \ref{sec:computational_problem}.
First, let us generate a path of spot prices $ S^{(0)}_{t}$ ($t \in \{ t_0, \ldots, t_n \})$ and temporarily view it as a ``real path'' observed in reality.
Suppose, at time $t = t_j$, the secondary trader wants to buy an option $S^{(i)}$, so it sends a query to the $i$-th primary trader about the current price.
Given the query, the primary trader simulates hypothetical future paths to learn the optimal hedging strategy.
The real path has already been observed up to time $t = t_j$, and the path branches off into a large enough number of simulations to ensure the convergence of training.
The left panel of Figure \ref{fig:sim-path} depicts this situation.
Since the above procedure is repeated for every ``real path'' and every time steps, the total number of hypothetical paths can be extremely large.

We now introduce our simulation design to sidestep this problem, which is illustrated in the right panel of Figure \ref{fig:sim-path}.
In our method, the path simulation for primary trader training does not branch off, and thus we use only one future path (per one real path) to train the primary trader.
In particular, we do not have to generate extra hypothetical paths for the primary trader training, since we can just reuse the real paths.
This is justifiable because our goal is to optimize the objective \eqref{eq:second-objective}, where we only need approximate gradients of the objective.
By this method, we can significantly reduce the number of simulations during training, while we can also increase the number of hedging situations (i.e., real paths) in training samples.

Note that our method resembles to the Least Square Monte Carlo (LSM) algorithm \cite{Longstaff2001} in the sense that it approximates expectation with a single simulated path.


\section{Experiments}
We conducted a series of experiments to demonstrate the efficacy of our proposed learning scheme. Here, we provide a common setup for our experiments.

As a hedging target $S^{(\star)}$, we considered an European call option with strike $= 1.00$.
Note that, in practical hedging scenarios, it is natural to consider more complicated options such as exotics.
For our purpose, however, it is enough to consider a vanilla European-type option because our experiments aim to compare the hedging performance of the proposed method to other baselines.
As hedging instruments, we employed the followings:
\begin{itemize}
  \item $S^{(0)}$: The underlying stock (explained in section \ref{sec:heston}).
  \item $S^{(1)}$: European Call Option (strike $= 1.02$).
  \item $S^{(2)}$: European Put Option (strike $=0.98$).
  
\end{itemize}

Other parameters are set as the following:
\begin{itemize}
  \item The steps to the maturity: $n=20$
  \item Parameters for the Heston model\footnote{Although the parameters are the same as the settings of original deep hedging \cite{deep-hedging}, the only difference is that $\sigma$ is set to enough smaller value that the Heston model fulfills Feller condition to avoid negative volatility.}: $S^{(0)}_{t_0}=1.0$, $V_{t_0}=0.04$, $\kappa=1.0$, $\theta=0.04$, $\rho=-0.7$, and $\sigma=0.3$
  \item Proportional cost coefficient for the underlying stock: \\$c_0 = 0.01, 0.005, 0.001, 0.0005, 0.0001$
  \item Proportional cost coefficient for the tradable options: \\$c_i = 0.01, 0.005, 0.001, 0.0005, 0.0001$
  \item The architecture of NN: 4-layered MLP with attention. (1 linear layer of 32-dim embedding, 2-layered self attention network, and 1 linear layer for outputs) All layers except for the last layer employs Layer Normalization \cite{ba2016} and ReLU Activation.
  \item $I_t$ (Information input to NN):
        \begin{itemize}
          \item log moneyness of all options
          \item time to maturity
          \item all trading prices of tradable assets
          \item the volatility of underlying stock
        \end{itemize}
  \item Utility function: depending on the experiments (explained later)
  \item Simulation paths of primary traders' training
  : 50,000
  \item The learning epochs of primary trader
  : 1,000
  \item Simulation paths of primary trader when the secondary trader is trained
  : 1,000
  \item Simulation paths of the secondary trader's training: 5,000
  \item Simulation paths of the secondary trader's test: 5,000
  \item The training epochs of the secondary trader
  : 500
  \item Optimizer of NN: Adam (learning rate$=10^{-3}$)
\end{itemize}

For the implementation, we used PFhedge\footnote{\url{https://github.com/pfnet-research/pfhedge}}.


\subsection{Experiment 1: Arbitrage Existence Test}
Firstly, we conducted an experiment to compare two types of primary traders: deep hedging and the Black--Scholes formula.
Specifically, we tested if arbitrage opportunities exist.
Since the Black--Scholes model is not consistent with the assumption of discrete-time market with frictions, it may offer arbitrage opportunities that is exploitable by the secondary trader.
To examine this possibility, we compared the performance of the above two types of primary traders.
If arbitrage opportunities are found in the option market, the secondary trader may overfit to such opportunities.
Hence, we may conclude that arbitrage opportunities do exist in the market if we observe that the Black--Scholes model performs unreasonably better than deep hedging.

For the objective of the secondary traders, we used the entropic risk measure (ERM), where the risk aversion coefficient is set as $\lambda = 1.0$ (see Equation (3.4) in \cite{deep-hedging} for a precise definition).
The reason for this choice is as follows:
In comparison with CVaR, ERM can be more sensitive to the expected profits because CVaR is only sensitive to left tails of profit distributions.
We expect that the effect of arbitrage opportunities would be more significant for ERM, hence the choice.

\subsection{Experiment 2: Hedging Performance Test}
Secondly, we compared risk reduction abilities of (i) the stock-only deep hedging and (ii) the nested deep hedging.
Since the stock-only deep hedging tries to hedge the target option $S^{(\star)}$ based solely on the underlying stock $S^{(0)}$, we expect that it fails to manage higher-order sensitivities such as Gamma and Vega.
We can corroborate this by showing that the nested deep hedging performs better than the stock-only method.
To understand the qualitative behaviors of hedging, we also examined values of several greeks.

We employed CVaR(0.1) as the objective function in this experiment.


\section{Results}
\subsection{Result of Experiment 1}

\begin{table}[tbp]
  \centering
  \caption{Hedge Result of Secondary Traders in Experiment 1.}
  \label{tab:results1}
  \begin{tabular}{c|ccc}
    \toprule
    Setting & \multicolumn{3}{c}{ERM-based Hedge Cost}\\
    Proportional cost ($c_0, c_i$) & Black--Scholes & Stock--only & Proposed \\ \midrule
    0.0001 (0.01\%) & 0.017436 & 0.022486 & 0.022510\\
    0.0005 (0.05\%) & 0.018976 & 0.022666 & 0.022508\\
    0.001 (0.1\%) & 0.018948 & 0.022789 & 0.022510\\
    0.005 (0.5\%) & 0.020064 & 0.022903 & 0.022665\\
    0.01 (1\%) & 0.021726 & 0.022986 & 0.022791\\\bottomrule
  \end{tabular}
\end{table}

Table \ref{tab:results1} shows the result of Experiment 1.
In the table, we show the performance of the following three methods:
\begin{itemize}
  \item Black--Scholes: The performance of the secondary trader when the primary trader employs the Black--Scholes model.
  \item Stock-only: The performance of the stock-only second trader, which is only using stock for hedging.
  \item Proposed: The performance of the second trader when the primary trader employs deep hedging.
\end{itemize}
The values in the table are the hedging cost of ERM, i.e., $u^{-1}(\mathbbm{E}[u(PL)])$.

According to the results, in the case of Black--Scholes, the hedging cost of the secondary trader is significantly small.
It means that the pricing of the Black--Scholes-based primary trader is inappropriate and gives an arbitrage chance to the secondary trader.
Although the chance becomes smaller when the transaction cost by the proportional cost is greater, arbitrage chance still seems to exist even when the proportional cost is 1\%.

\subsection{Result of Experiment 2}
\begin{table}[tbp]
  \centering
  \caption{Hedge Result of Secondary Traders in Experiment 2.}
  \label{tab:results2}
  \begin{tabular}{c|cc}
    \toprule
    Setting & \multicolumn{2}{c}{CVaR(0.1)-based Hedge Cost}\\
    Proportional cost ($c_0, c_i$) & Stock--only & Proposed \\ \midrule
    0.0001 (0.01\%) & 0.031036 & 0.025988\\
    0.0005 (0.05\%) & 0.031792 & 0.026323\\
    0.001 (0.1\%) & 0.032624 & 0.027061\\
    0.005 (0.5\%) & 0.038762 & 0.028480\\
    0.01 (1\%) & 0.045453 & 0.031614\\\bottomrule
  \end{tabular}
\end{table}


Table \ref{tab:results2} shows the results of secondary traders' hedging.
In this table, values are the hedging cost defined from CVaR(0.1).
According to the results, the proposed secondary trader successfully reduces the hedging costs.

\begin{figure}[tb]
  \centering
  \includegraphics[width=0.6\linewidth]{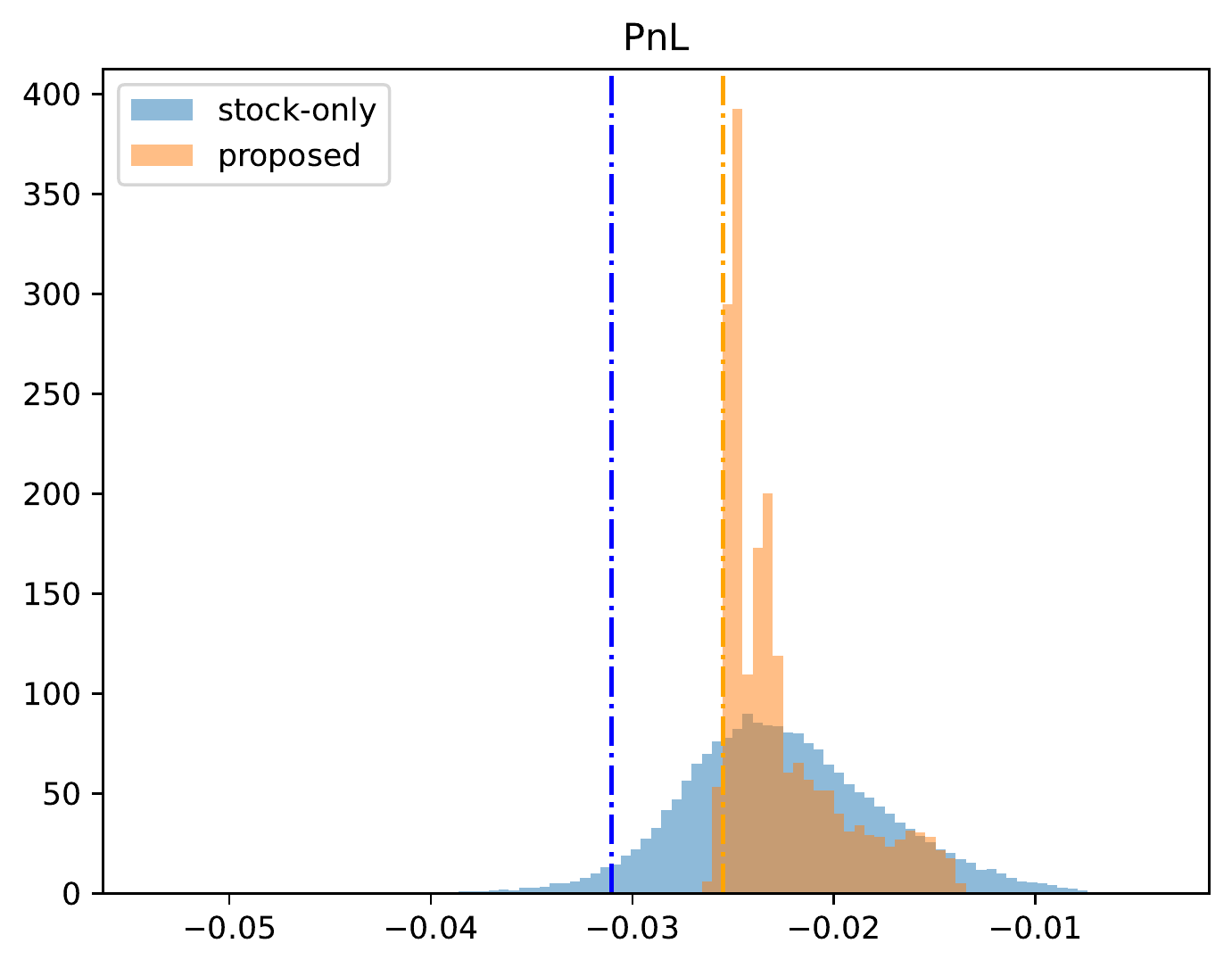}
  \vspace{-3mm}
  \caption{PL of secondary traders. (The proportional cost is 0.0001) The broken lines mean CVaR(0.1).}
  \label{fig:cvar}
  \centering
  \includegraphics[width=0.49\linewidth]{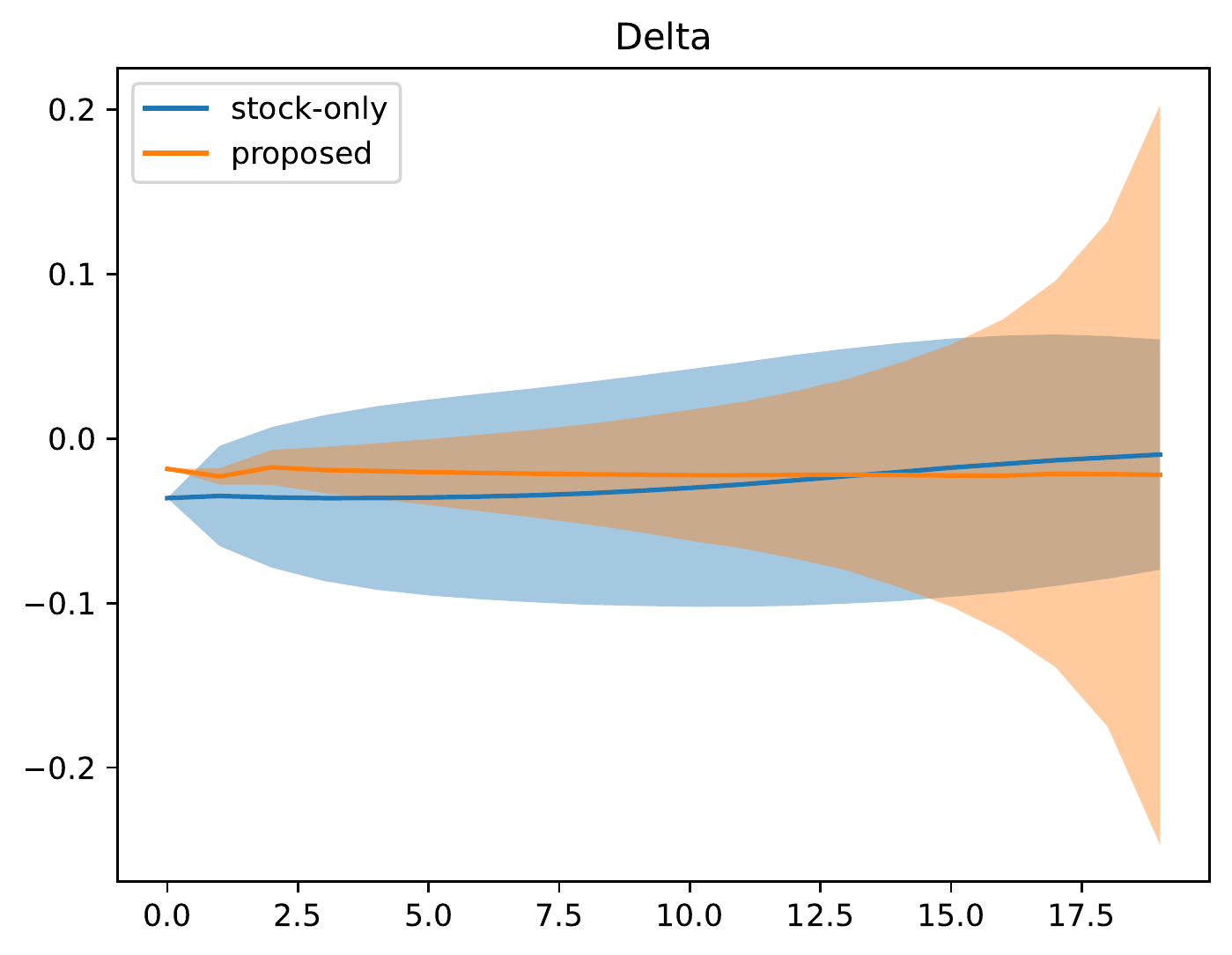}
  \includegraphics[width=0.49\linewidth]{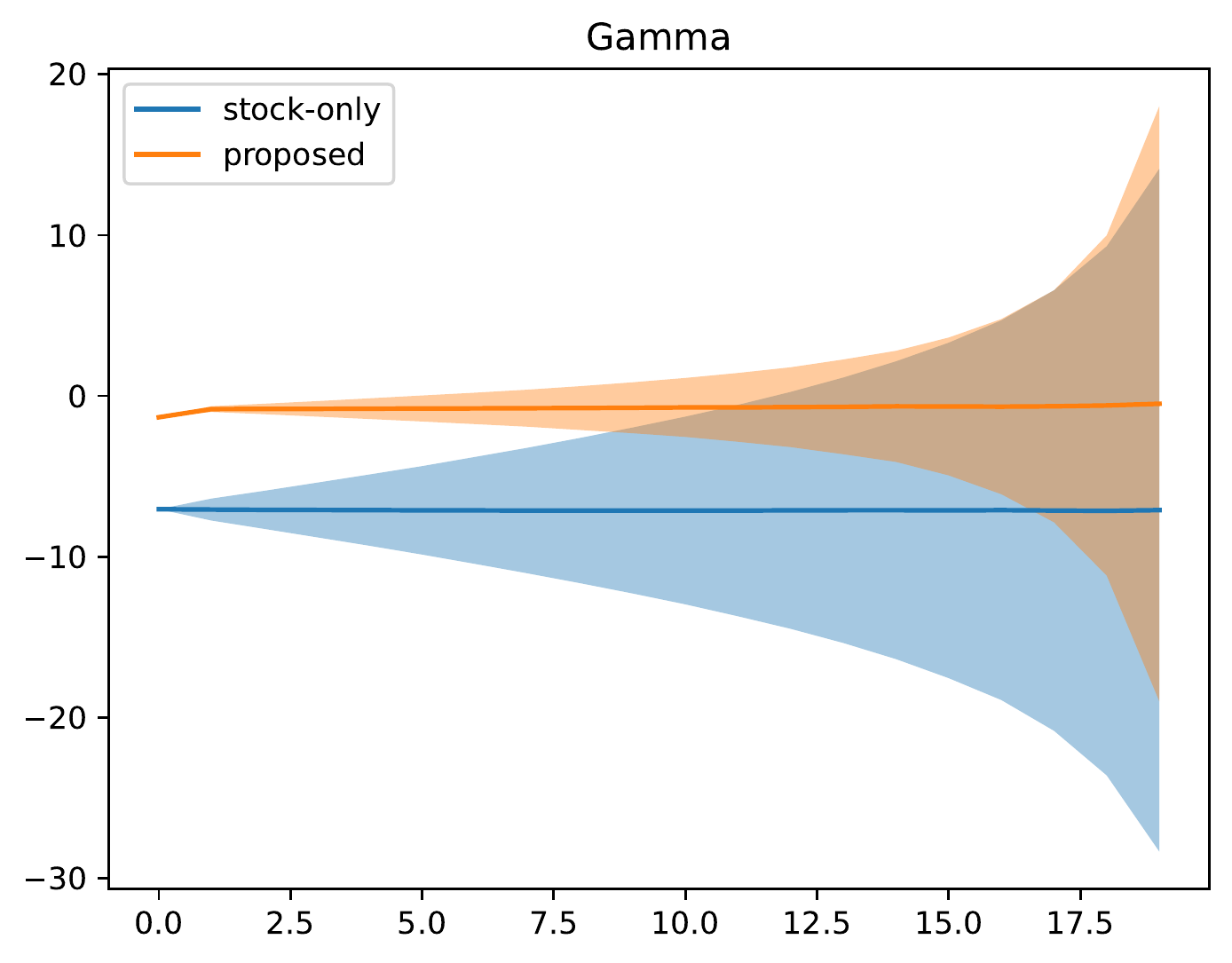}
  \includegraphics[width=0.49\linewidth]{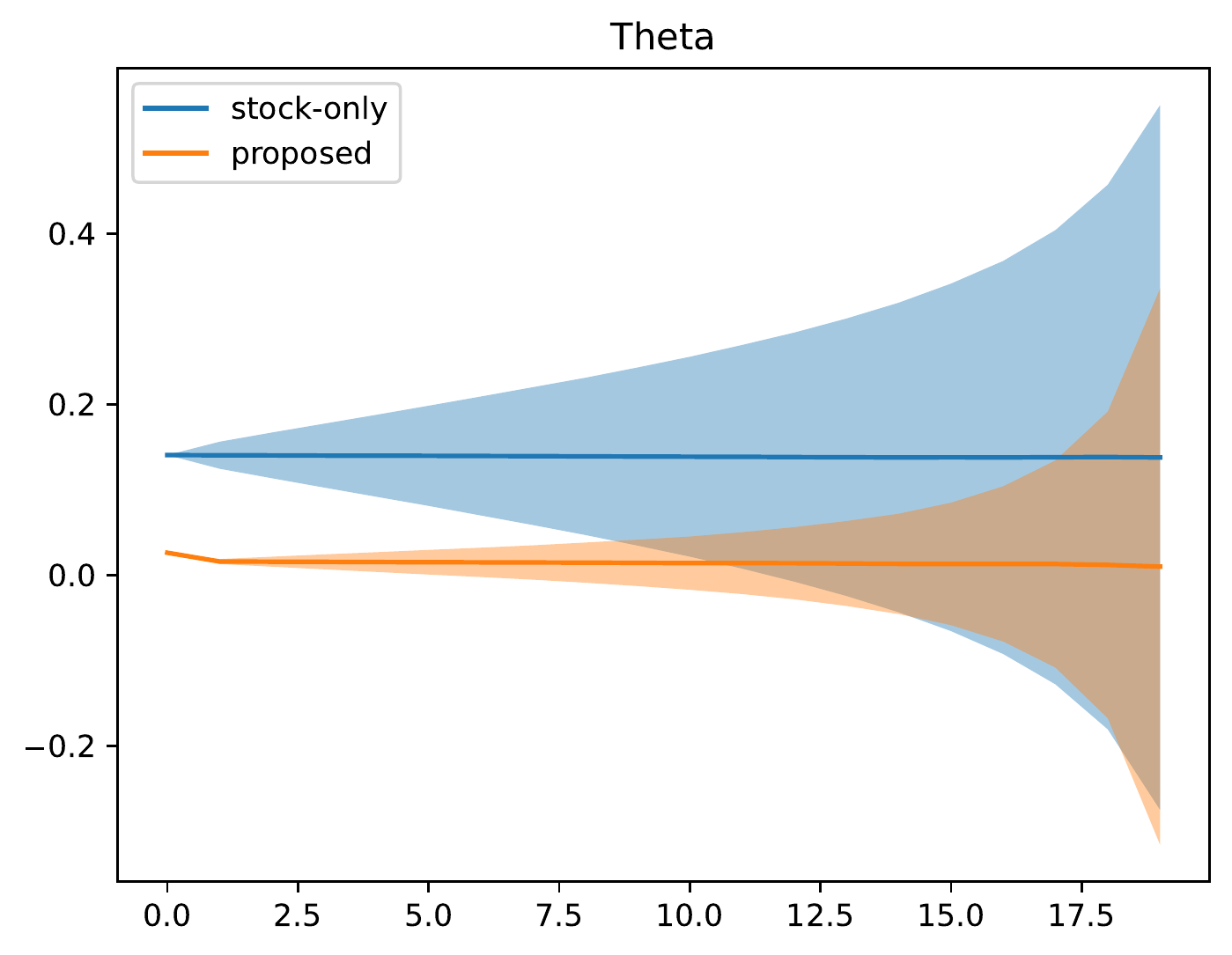}
  \includegraphics[width=0.49\linewidth]{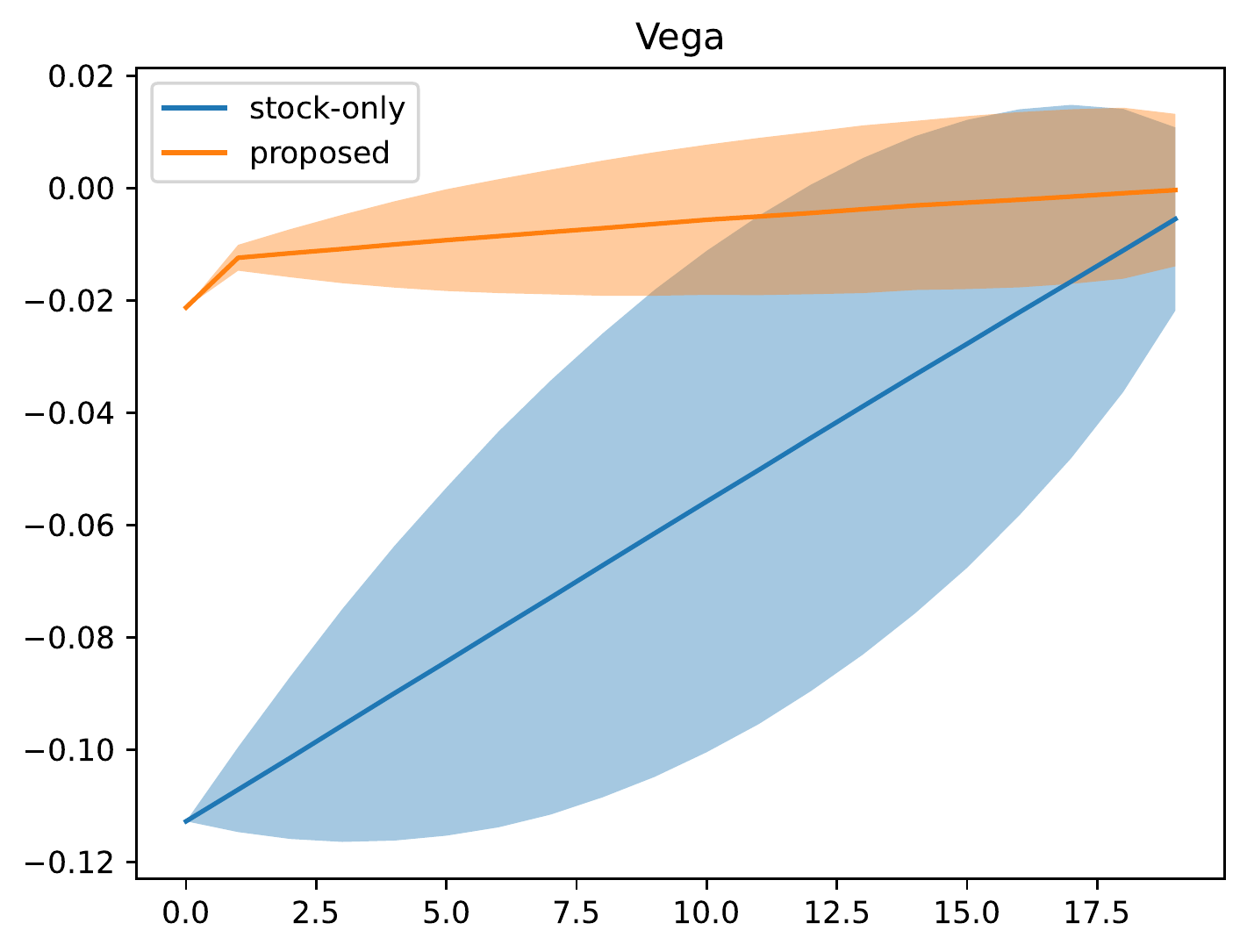}
  \vspace{-5mm}
  \caption{Greeks of the portfolios of secondary traders. (The proportional cost is 0.0001) A horizontal axis means that time steps. The filled areas are 95\% CI.}
  \label{fig:greeks}
\end{figure}

Figure \ref{fig:cvar} plots the distribution of PL (higher is better).
Broken lines indicate the CVaR(0.1) of each distribution.
In this figure, PL is used instead of hedging cost.
According to this plot, the proposed secondary trader successfully reduces only the tail risk, which effectively reduces CVaR(0.1).

Lastly, we also plot several greeks of hedging portfolio to investigate how the secondary traders work.
Figure \ref{fig:greeks} shows the results of greeks (delta, gamma, theta, and vega).
The proposed secondary trader successfully reduces the greeks except for delta more than the stock-only secondary traders.
Interestingly, the stock-only secondary trader seems to outperform the proposed model in terms of the delta.

\section{Discussion}

First, according to Experiment 1, the proposed secondary trader has no significant difference in ERM from the spot-only secondary trader, which suggests that no significant arbitrage chance exists and the pricing by the deep-hedging primary trades is appropriate.
On the contrary, the Black--Scholes secondary trader shows a significant cost reduction, suggesting that arbitrage chance does exists in the pricing by Black--Scholes-based primary traders.
This result supports our motivation that the prices of tradable options as hedging instruments should be calculated by deep-hedging instead of the Black--Scholes model to obtain a better strategy for the secondary trader.
It proves the rationality of our proposed learning methods.

Second, according to Figure \ref{fig:cvar}, the proposed secondary trader successfully improves CVaR(0.1).
It is anticipated from the fact that the proposed secondary trader can also utilize tradable options for hedging, while the stock-only secondary trader cannot.

In terms of greeks, the proposed secondary trader succeeds better in greek-free.
This result also supports that our proposed learning method works properly.
Interestingly, stock-only secondary traders perform slightly better than the proposed method, while they seem to fail in other types of greeks.
From these observations, we may conclude that the stock-only secondary trader performs better in the delta because the learning for the stock-only secondary trader can focus only on the delta, unlike the proposed secondary trader that is capable to control multiple variables.

It is worth noticing that the improvement in the average profit is quite subtle, while the proposed method does reduce the risk measure.
This might be because the nested deep hedging does not admit significant arbitrage opportunities.
As we observed from Experiment 1 and Figure \ref{fig:cvar}, it seems hard to obtain additional profits even by the proposed secondary trader.

Again, we emphasize that our proposed learning method is critical for the effective nested deep hedging, because otherwise it could be computationally intractable.

We believe that our work is a step towards real-world applications of deep hedging, but several issues are left for future work.
For example, liquidity have to be considered.
In this study, we have assumed that the hedge instruments are sufficiently liquid so that the primary trader is completely under the control of the secondary trader.
However, we should consider liquidity of instruments under more realistic settings.
Other directions include generalizing our results to other types of frictions.
We have already investigated the case of the significantly high proportional cost, but other models of frictions are worth investigating.
Also, developing efficient learning method for exotic options is interesting future work.

\section{Conclusion}
We considered the problem of hedging an option with multiple options.
In this setting, options used as hedge instruments have to be priced during the trading period.
To develop a deep hedging approach to this setting, it is sensible to coherently employ deep hedging for both the hedged and tradable options, in order to avoid exploitable arbitrage opportunities.
However, the overall training process can be computationally intractable when we straightforwardly apply the existing learning scheme to this setting.
To overcome this issue, we proposed a novel learning technique for  \textit{nested} deep hedging, which allows us to hedge an option with options while circumventing computational intractability.
We conducted a series of experiments to investigate the existence of arbitrage chances and the performance of our proposed method.
We showed that arbitrage chances can arise when the classical Black--Scholes pricing formula is used for hedging instruments, while the coherent use of deep hedging eliminates such arbitrage chances.
Moreover, we demonstrated that our proposed method successfully reduced tail risks and exposures to greeks.
We would like to stress that the above results are made possible by our proposed learning method, because otherwise the overall computation in the experiments could be computationally intractable.
Future research can address a more realistic market setting and extend our result to exotic options.

\section*{Acknowledgment}
We would like to thank Yasuhiro Fujita for reviewing our paper and giving us valuable comments.

\bibliographystyle{IEEEtranS}
\bibliography{cite}

\end{document}